\documentclass[preprint]{elsarticle}
\makeatletter
\def\ps@pprintTitle{%
	\let\@oddhead\@empty
	\let\@evenhead\@empty
	\let\@oddfoot\@empty
	\let\@evenfoot\@oddfoot
}
\makeatother
\usepackage{lineno,hyperref}
\modulolinenumbers[5]
\usepackage{setspace}
\journal{Physics Letters B}

\usepackage{graphicx}
\usepackage{fleqn}
\usepackage{amsfonts}
\usepackage{amsmath,changepage}
\usepackage{amssymb}
\usepackage{braket}
\DeclareMathAlphabet{\mathpzc}{OT1}{pzc}{m}{it}

\begin{document}
 	\title{Planar diagrams for lattice gauge theory on finite tori.}
 	\author{Herbert Neuberger}\fnref{myemail} \fntext[myemail]{Email: herbert.neuberger@gmail.com}
 	
 	\address[add1]{Present Address: Department of Particle Physics and Astrophysics,
 		Weizmann Institute of Science,Rehovot 7610001, Israel.}
 	
 	\address[add2]
 	{Permanent Address: Department of Physics and Astronomy, Rutgers University,  Piscatway, NJ 08854, U.S.A.}

	\begin{abstract}
    An $N=\infty$ equivalence among quenched $U(N)$ models on finite lattice tori of $V$ sites is proven to all orders in planar perturbation theory by putting circulant lattice momenta together with group indices on 't Hooft's double lines. Known estimates for the number of order $N^2$ diagrams, $N\gg V$, and the simultaneous presence of UV and IR cutoffs, suggest a positive radius of convergence for the planar perturbative expansion before the limit $N=\infty$ is taken.
    
	\end{abstract}
	\maketitle
	\section{Introduction.}
	The large $N$ expansion in gauge theory originated in~\cite{tHooft:1973alw}. Eguchi and Kawai (EK)~\cite{Eguchi:1982nm} proved that  $\lim_{N\to\infty} U(N)$  4-dimensional lattice gauge 
	theory can be reduced to a 4-matrix model in the strong coupling regime, but it was soon found to fail in weak coupling perturbation theory. The Quenched Eguchi-Kawai (QEK )~\cite{Bhanot:1982sh} model was proposed as an extension of EK to weak couplings, and shown to reproduce the lattice Coulomb law, without affecting the strong coupling regime. Subsequently, Parisi~\cite{Parisi:1982gp} applied quenching to a  non-gauge matrix model where it held to all orders. Starting from that observation, Gross and Kitazawa ~\cite{Gross:1982at} investigated quenched gauge theory at all orders with various UV cutoffs, including the lattice.
	
	The EK strong coupling proof holds on a hypercubic lattice of arbitrary finite volume at $N=\infty$. In this work, validity of quenched weak coupling perturbation theory is established to all orders in the same set of models.  Peter Orland ~\cite{Orland:1983gt} (up to equation (4a) therein) anticipated the starting point of this article, but continued differently. 
	
	\section{Preliminary.}\label{prelim}

	\subsection{Elementary lattice transporters.}	
	This work is about lattice $U(N)$ gauge theories with Wilson action on a $d$ dimensional rectangular torus of sides $L_\mu , \mu=1,..,d;\; 1\le L_\mu < \infty$ with  $V=\prod_{\mu=1}^d L_\mu$ sites.  $U(N)\supset SU(N)$ is used for simplicity. Dynamical variables are labeled by site and group indices taking $M=VN$ values. Also for  simplicity, $N,L_1,..., L_d$ are assumed pairwise co-prime, $N > L_1 \ge L_2....\ge L_d$. $M$ determines the ranges of group and site indices; $Z(M)\cong Z(N)\otimes Z(V)\cong Z(N)\otimes_\mu Z(L_\mu)$.  
	
	Unit matrices acting trivially in various factor spaces will be suppressed. For fixed $\mu$, the $M\times M$ unitary matrix $T_\mu$
	generates minimal distance parallel transport in the $\pm\mu$ directions. Ordered products of $T_\mu$, $T^\dagger_\mu$ with various $\mu$, act on site anchored ancillary fields $\{\phi_x^j\}$ in the  $U(N)$--fundamental representation and define parallel transport along all lattice paths. Link matrices $U^{ij}_{x,\mu}\in U(N)$ have a $d$-component site index $x$ ($x_\mu=0,..,L_\mu-1$), and $i,j=1,..,N$ $U(N)$ group indices. $x^{\perp (\mu)}$ has $x^{\perp (\mu)}_\mu =0$, denoting sites on a ($d-1$)--hyperplane orthogonal  to the $\mu$ direction. Every site $x$ decomposes as $x=x^{\perp(\mu)}+x^{\parallel (\mu)}$ with $x^{\parallel (\mu)}_\mu \in \{0,1,..,L_\mu -1\}$ and $x^{\parallel (\mu)}_\nu =0,\;\forall\;\nu\ne\mu$. $T_\mu, T_\mu^\dagger$ are  diagonal in $x^{\perp(\mu)}$. They only connect $x^{\parallel (\mu)}_\mu$ to $x^{\parallel (\mu)}_\mu \pm 1$ mod-$L_\mu$. The $T_\mu\in U(N)$, $1\le \mu\le d$, are subjected to constraints setting some entries to zero: locality is reflected by sparseness for $V>1$. All $V$ admit multi-indices replacing  EK group indices (upper case) by $d+1$ component ones: the first $d$ are collected into a $d$-vector $x$ and the last one is a (lower case) group index. The indices take $M=NV$ distinct values.
	\begin{equation}\label{eqn:tmu}
	[T_\mu^0 \phi ]_x^i=\phi_{x-\mu}^i\;\; 
		T_\mu = T_\mu^0 \mathbf{U^\mu}, {\rm where}\;[\mathbf{U^\mu}\phi]_x^i =\sum_j U_{x,\mu}^{ij}\phi_x^j .
	\end{equation} 
$T_\mu$ only contains $\mu$-link matrices. A gauge transformation by $g_x\in U(N)$, $[^{g}\mathfrak{G}\phi]_x=g_x\phi_x$, acts on $T_\mu$ by conjugation:
	\begin{equation}
		[ (^{g}\mathfrak{G})^\dagger T_\mu({\mathbf U}_\mu) ( ^{g}\mathfrak{G} ) \phi ]_x =T_\mu (g_x^\dagger U_{x-\mu , \mu} g_{x-\mu}\phi_{x-\mu})\equiv [T_\mu ( ^{g}{\mathbf U}_\mu) \phi]_x .
	\end{equation}
	The partition function with Wilson action is: 
	\begin{equation}
		\label{eqn:Z}
		Z_W=\int [\prod_{x,\mu} dU_{x,\mu}] e^{-N\beta \sum_{\mu < \nu} \lVert [T_\mu,T_\nu]\rVert^2},\;\;\;\lVert X \rVert^2 := {\rm Tr} X X^\dagger.
	\end{equation}
	$\beta$ is 't Hooft's lattice coupling, $dU_{x,\mu}$ is Haar and the action is gauge invariant. It also has a global center symmetry, $U(1)^d$ under  ${\mathbf U}_\mu \rightarrow
	 e^{i\chi_\mu} {\mathbf U}_\mu$ with $0\le\chi_\mu < 2\pi$,  $T_\mu \leftrightarrow T_\mu^\dagger$ conjugation symmetries and $T_\nu \rightarrow {T^0_\mu}^\dagger T_\nu T^0_\mu\;\forall\nu$ cyclic translation symmetry in $\mu$. The group of translations is $Z(V)$.
	 
	\subsection{Discrete Fourier Transform (DFT).}
	Expansion in $\beta^{-1}$ simplifies by application of $DFT$ to gauge field perturbations. For one direction, ignoring the label $\mu$ and gauge group indices, the DFT of $T^0$ is ${\hat T}^0=\Omega  T^0 \Omega^\dagger$ where,
	\begin{equation}
		\Omega^{pq} = \frac{1}{\sqrt{L}} \omega^{-pq},\;\;\; p,q=0,1...,L-1, \;\;\; \omega =e^{-\frac{2\pi i}{L}}, \;\;\; \Omega^\dagger=\Omega^*=\Omega^{-1},
	\end{equation}
	with arithmetic mod--$L$ on indices $p$ (momentum) and $q$ (site) labels. 
	\begin{equation}
		\begin{split}
		&T^0 \Omega=\Omega \; {\rm diag}(1,\omega,..,\omega^{L-1}),\;\; T^0 = \Omega \;{\rm diag}(1,\omega,..,\omega^{L-1})\Omega^\dagger ,\\&
		(\Omega v )_p = {\hat v}_p =\frac{1}{\sqrt{L}} \sum_q \omega^{-pq} v_q ,\;\; (\Omega^\dagger {\hat v})_q=v_q=\frac{1}{\sqrt{L}} \sum_p \omega^{qp} {\hat v}_p ,\\&
		[\Omega \;{\rm diag}(v_0, v_1,...,v_{L-1})\; \Omega^\dagger]_{pp'} =\frac{1}{L} \sum_q \omega^{-pq}  v_q\; \omega^{qp'}={\hat v}_{p-p'} ,\;\;\sum_q v_q=L{\hat v}_0.
		\end{split}
	\end{equation}
	Summation over $p$ and $p'$ of a quantity depending only on $r=p-p'$ gives a factor of $L$ times a summation over $r\equiv p-p'$ in the range $0$ to $L-1$. 
	When the $v_q$ are independent complex numbers, so are the ${\hat v}_p$. 
	The $L\times L$ matrix $\hat V$, $({\hat V})_{pp'}\equiv{\hat v}_{p-p'}$ is circulant, ${\hat V} = {\rm circ}({\hat v}_0, {\hat v}_1,....{\hat v}_{L-1})$; 
	if ${\hat V}$ is diagonal, ${\hat V} = {\hat v}_0 {\mathbf 1}$.  ${\rm Tr} {\hat V}{\hat V}^\dagger = \sum_q |v_q|^2 = \sum_{p,p'} |{\hat v}_{p-p'}|^2=L\sum_p|{\hat v}_p|^2$.

	Circulant matrices close under matrix product and ensure virtual particle momentum conservation in diagrams. 
    The indices $p$ and $p'$, live on oppositely oriented lines of 't Hooft's propagators which carry total momentum $r$. 
	
	The complete DFT on a $d$-dimensional lattice is $\Omega=\prod_\mu^{d}\Omega_\mu$
	each factor operating on $\mu$-indices with mod-$L_\mu$ arithmetic on $q_\mu , p_\mu$. 
	The notations $x^{\perp(\mu)},  x^{\parallel(\mu)}$ extend to their conjugate momenta $p^{\perp(\mu)}, p^{\parallel(\mu)}$. For each fixed pair of group indices
	the gauge field is circulant in momentum space. $T_\mu\rightarrow{\hat T_\mu}:=\Omega^\dagger T_\mu \Omega$ leaves
	the action in~(\ref{eqn:Z}) unchanged. In addition to a circulant structure, the site sparseness properties of  $T_\mu$ in momentum space become:
	\begin{equation}\label{unidir}
	{[{\hat T}_\mu]}_{{(p^{\perp(\mu)},p^{'{\perp(\mu)}})}} =\delta_{(p^{\perp(\mu)},p^{'{\perp(\mu)}})},
	\end{equation}
	where $p^{\parallel(\mu)}$ and group indices are suppressed.
    
	\section{Perturbation theory for $U(N)$ on lattice tori of $V<\infty$ sites (EKV).}\label{unq-torii}
	
	EK has a single site, so ${\hat T}^{EK}_\mu \equiv {T}^{EK}_\mu$. It is convenient to stay in momentum space for all $V$.
	Perturbation theory consists of summing up 
	contributions associated with distinct absolute classical minima
	and independent expansions about them.

	\subsection{All absolute minima of EKV.}
	
	We start by temporarily ignoring sparseness constraints on ${\hat T}_\mu$.

	 In EKV we need $[{\hat T}_\mu , {\hat T}_\nu]=0,\;\;\forall \mu,\nu$. Hence, ${\hat T}_\mu = {\hat D}_\mu, \forall \mu$:
	\begin{equation}\label{eqn:mesh}
		[{\hat D}_\mu ]_{(pj),(p'j')}=\prod_{\nu\ne\mu}^d (\delta_{p_\nu,p'_\nu}) \delta_{j,j'} d^\mu_{p} d^\mu_{(pj)},\;\;\;d^\mu_{p}=e^{2\pi i p_\mu / L_\mu},\;\;\;
		d^\mu_{(pj)}=e^{i\vartheta^\mu_j /L_\mu}.
	\end{equation} 
    The first factor, $d^\mu_p$ in ${\hat T}_\mu$ is purely diagonal 
    while the second is circulant in $p$, unitary and diagonal. Consequently, there is no explicit dependence on $p$ in the second factor. Moreover, $d^\mu_p$ does not depend on $p^{\perp(\mu)}$, rendering each $\vartheta^\mu_j$  $\frac{V}{L_\mu}$--fold degenerate.  The division by $L_\mu$ ensures that the dependence is on $N$ angles all within a $2\pi$ range which together make up a gauge invariant set. We end up with $M$ nonzero entries in the diagonal matrices with distinct $\vartheta^\mu_j$ values coming in clusters of  $\frac{V}{L_\mu}$ degenerate angles, distinguished only by $p^{\perp(\mu)}$ indices. There is one such cluster for each $p_\mu$. One value for $\vartheta^\mu_j$ appears $V$ times on the diagonal of ${\hat D}_\mu$. 
	
	In site-space the minima are characterized by loops winding once and going straight in the $\mu$-direction having the same eigenvalues for each $\mu$, but away from minima the angles become $x^{\perp(\mu)}$ dependent. The integration has Haar factors, $\mathfrak{H}(\mathbf{\vartheta^\mu}  )$,  for each $\mu$-directed once winding loop.
	\begin{equation}\label{eqn:haar}      	
		\prod_\mu \prod_{x^{\perp(\mu)}}\left\{	\mathbf{\mathfrak{H}} (\vartheta^\mu (x^{\perp(\mu)}))  [ \int\prod_{l=1}^N d {\vartheta^\mu}_l(x^{\perp(\mu)}) ]  \right\}.
	\end{equation}
	At the minima, products over $x^{\perp(\mu)}$ get replaced by powers $V/{L_\mu}$ of $\mu$--Haar factors. In contrast to $V=1$, one cannot explicitly separate integration over eigenvalues and cosets at all orders. This fact is irrelevant to perturbation theory. The Haar measure ensures that quenched angles with identical $d$-momenta would 
	spread uniformly. Without quenching the distribution could have peaks  as a result of an attraction induced by feedback from Gaussian fluctuations.

	 \subsection{Bijection between EK and EKV indices.}
	 
	  We always have $M$ dimensional bases. The EK--EKV difference is that momenta are ``naturally'' quenched while the more general quenching of group indexed angles in QEK~\cite{Bhanot:1982sh} is by decree. Equation (\ref{eqn:mesh}) shows that momenta and angles mesh.  
	 
	  EK $J,J'...$ and EKV indices $(p,j),(p',j'),...$ can be related by an invertible relabeling determined by the angles:  For each $\mu$ the $\mu^{\rm th}$--$S_\mu \subset S^d=\prod_\mu S_\mu$, the momenta torus, is divided into consecutive $\frac{2\pi}{L_\mu}$ arcs. It is assumed that we are given $N$ angles $0\le\vartheta^\mu_j <2\pi , j=1,...,N$ non-decreasing as $j$ increases. Consecutive arcs, ordered by increasing values of the starting point $p_\mu = 0, 1, 2, . . L_\mu-1$, are populated by angles of ascending values $ [2\pi p_\mu +\vartheta^\mu_j ]/ L_\mu,\;\; j=1,...,N$. After this is done for each $\mu$ we have  $N\prod_\mu L_\mu = M$ points on the circle $S_\mu$ and we can label them by $J=1,..,M$ starting with $\mu=1$, going around it in ascending $J$-order and continuing to the next $\mu$ until $\mu=d$ is finished. The result is a one-to-one correspondence $J_{(p,j)}$ with inverse $(p,j)_J\equiv (p_J , j_J )$.
	EKV analogues of EK equations are obtained by a replacement of the respective minima angles: 
	\begin{equation}\label{eqn:repl}
	\vartheta^\mu_{J_{(p,j)}}\rightarrow [2\pi (p_J)_\mu + {\vartheta_{j_{J}}}^\mu] / {L_\mu}.  
	\end{equation}

    \subsection{Review of EK and quenching.}\label{ekrev}
    The form of equation (\ref{eqn:Z}) is independent of $V$. Using the $J$ to $(p,j)$ bijection perturbation theory in quenched EK gives, {\it  mutatis mutandis} the one in EKV, where $V>1$ is included. Below is a brief review of the EK calculation in ~\cite{Bhanot:1982sh} where $V=1$, there are no sparseness restrictions and it is possible to separate the integration variables into $\vartheta_\mu^J$ sets and their cosets. The absolute minima of the action are given by commuting $T^{EK}_\mu$  matrices.  When $\beta$ becomes very large and $M\to\infty$, $V=\infty$ lattice Feynman diagrams cannot be recovered, but the strong coupling expansion is, as a consequence of the preservation of $U(1)^d$ ~\cite{Eguchi:1982nm}. 
     
    The main point of ~\cite{Bhanot:1982sh} was that the $\vartheta_\mu^J$ of $T^{EK}_\mu$ act as the compact $\mu$-component of lattice momenta at tree level. The momentum $d$-dimensional torus is flat and the dependence on momenta trigonometric. Quenching is needed in order to preserve this at $M\to\infty$. It maintains the validity of unquenched EK at strong coupling and forces global center $U(1)^d$ invariance at weak coupling.

    The minima $T^{EK}_\mu = D_\mu:={\rm {diag}}(e^{i\vartheta^\mu_J} ; J=1,..M)$ come with a Haar measure factor $\mathfrak{H}(\vartheta^\mu)$ for each $\mu$, 
    \begin{equation}\label{eqn:haarEK}
    \mathfrak{H}(\mathbf{\varphi}  ):=\prod_{J > J'} \sin^2 ( \frac{\varphi_J -\varphi_{J'}}{2}).
    \end{equation}
    The superscript EK is dropped from here to end of this section~(\ref{ekrev}).
    At order $n$ up to rank $n$ polynomials in $A_\mu^{JJ'}$ need to be kept in $T_\mu$:
    \begin{equation}\label{eqn:EK1}
    		T_\mu = e^{i A_\mu} D_\mu (\vartheta^\mu_J) e^{-i A_\mu}=
    		\sum_{n=0}^\infty \frac{i^n}{n!} \underbrace{[A_\mu,...,[ A_\mu}_n , D_\mu(\vartheta^\mu_J)\underbrace{]..]}_n \equiv e^{i\mathbf{A}\mathbf{d}(A_\mu)} D_\mu ,
    \end{equation}
    where $\mathbf{A}\mathbf{d}(A) D = [A,D]$. Integration over $A_\mu$ comes with Haar measure:
    \begin{equation}
    	|\det [E (A_\mu) ]|\;\; {\rm where}\;\; E(A_\mu):= \sum_{n=0}^\infty \frac{{[\mathbf{A}\mathbf{d}(A_\mu)]^n}}{(n+1)!}.
    	\end{equation}
    
    Fluctuations of the $\propto M$ angles are ignored as $M\to\infty$ because they are compact~\cite{Dothan:1979rj}. $[{\hat A}_\mu ]_{JJ}, J=1,..,M$ will be ignored, leaving only complex perturbative fields  ~\cite{tHooft:1973alw}. 
    
    The first order variation of the commutator ${\big [} T_\mu , T_\nu  {\big ] }_{J J'}$ is:
    \begin{equation}\label{eqn:EK2}
    i{\big (}[ A_\mu ]_{J J'}-[A_\nu ]_{J J'}{\big )}
    {\big (}[ D_\mu ]_{J'}(\vartheta^\mu_{J'})-[D_\mu ]_J(\vartheta^\mu_J ){\big )}\;
    {\big (} [D_\nu ]_{J'}(\vartheta^\nu_{J'})-[D_\nu ]_J(\vartheta^\nu_J ){\big )}.
    \end{equation}
    The contribution at second order to ${\rm Tr}{\big [} T_\mu , T_\nu {\big ]} {\big [} T_\mu , T_\nu {\big ]}^\dagger$ at fixed $\mu,\nu$ is:
    	\begin{equation}\label{eqn:EK3}
    		\begin{split}
    			 & \sum_{J\ne J'} {\big |}[A_\mu ]_{J J'}\!-\![ A_\nu ]_{JJ'}{\big |}^2{\big |}([D_\mu ]_{J'}(\vartheta^\mu_{J'})\!-\![D_\mu ]_J(\vartheta^\mu_J ){\big |}^2 
    			{\big |} [D_\nu ]_{J'}(\vartheta^{\nu}_{J'})\!\\&-\![D_\nu ]_j (\vartheta^\nu_J ){\big |}^2  =
    			16 \sum_{J\ne J'} {\big |}[A_\mu ]_{JJ'}\!-\![A_\nu ]_{JJ'}{\big |}^2\sin^2 \frac{ \vartheta^\mu_J\!-\!\vartheta^\mu_{J'}}{2} 
    			\sin^2 \frac{\vartheta^\nu_J \!-\!\vartheta^\nu_{J'}}{2}.
    		\end{split}  	
    	\end{equation}
    	The quadratic piece of the action is:
    	\begin{equation}\label{eqn:EK4}
    		\begin{split}
    			&\sum_{\mu, \nu} {\rm Tr}{\big [} T_\mu , T_\nu {\big ]} {\big [} T_\mu , T_\nu {\big ]}^\dagger  = 
    			32 \sum_{\mu,J\ne J'} {\big |}[A_\mu ]_{JJ'}{\big |}^2\sin^2 
    			\frac{\vartheta^\mu_J -\vartheta^\mu_{J'}}{2} 
    			\\\times&\sum_{\nu} \sin^2 
    			\frac{\vartheta^\nu_J -\vartheta^\nu_{J'}}{2}
    			-32\sum_{J\ne J'}{\bigg |} \sum_\mu [A_\mu ]_{JJ'}\sin^2 \frac{\vartheta^\mu_J -\vartheta^\mu_{J'}}{2}{\bigg |}^2. 		
    		\end{split}  	
    	\end{equation}
    The $A_\mu$-propagators are diagonalized in $\mu$-labels by adding $32\sum_{J\ne J'}\sum_\mu |{S_{\rm gf}}_\mu^{JJ'}|^2$ to the action with gauge fixing function ${S_{\rm gf}}_\mu=i[D_\mu , A_\mu ]$ and propagators now attain a Feynman--gauge structure. The Faddeev-Popov determinant, 
    \begin{equation}\label{eqn:EK5}
    	\prod_{J\ne J'} \sum_\mu \sin^2 \frac{\vartheta^\mu_J-\vartheta^\mu_{J'}}{2}.
    	\end{equation} 
    multiplies the existing Haar factors. 
    
    If  $\vartheta^\mu_J = \vartheta^\mu_{J'}$ with $J\ne J'$ the $[A_\mu ]_{JJ'}$ propagator is apparently divergent. The corresponding degrees of freedom are eliminated by the sparseness constraints at $V>1$. 
    
    To one loop, the probability distribution of $\vartheta^\mu_J$, $ P(\vartheta)$, obtained by integration over the complex $[A_\mu ]_{JJ'\; J\ne J'}$ exactly cancels the Haar measure and partially the Faddeev-Popov determinant, leaving
    \begin{equation}\label{eqn:EK6}
    	P(\vartheta ) \propto [\prod_{J\ne J'} \sum_\mu \sin^2 \frac{\vartheta^\mu_J-\vartheta^\mu_{J'}}{2}]^{2-d}.
    	\end{equation}
    For $d>2$ angle repulsion turns into attraction for well separated angles (where it is credible) and the $U(1)^d$ breaks as $N\to\infty$.   
    
    Quenching eliminates by decree the feedback of gauge field fluctuations on the distribution of angles. The \underline{expectation value} 
    of a gauge invariant observable is to be evaluated at fixed angles and only thereafter averaged over a distribution that becomes flat when the number of angles goes to $\infty$. The Haar measure does this and is a natural choice if one wishes to preserve strong lattice coupling expansion features. Other alternatives to the details of quenching are acceptable ~\cite{Bars:1983vw}. In perturbation theory only the region in gauge field space in the immediate neighborhood of the quenched angles is explored. For quenching to stay valid beyond perturbation theory in the planar limit, integration over cosets has to be fully implemented ~\cite{Neuberger:2020wpx}.
    
    Equations (\ref{eqn:EK1}-\ref{eqn:EK6}) determine the diagrammatic structure of EK to order $1/M$. We have a matrix model consisting of $d$ $M\times M$ hermitian matrices and a perturbative expansion with diagonal propagators in their non-diagonal entries. The vertices are defined by $dM$ angles contained in $d$ diagonal unitary matrices $D_\mu$ which enter linearly in each $T_\mu$ contribution. Wilson's action is a sum over all pairs $\mu,\nu$ of  $ \lVert [T_\mu,T_\nu]\rVert^2$. All absolute minima of the action have $T_\mu=D_\mu$ where the $D_\mu, \mu=1,..,d$ are simultaneously diagonal and the phases on their diagonals make up $d$ gauge invariant parameter sets. The propagators have simple expressions in terms of analytic periodic functions of phase differences.
	 
	 \subsection{EKV expansion in $\beta^{-1}$.}
	 
	 The perturbative expansion of the action contains traces of polynomials in ${\hat A}_\mu$ with up to four insertions of diagonals. Propagators connect
	 ${\hat A}_\mu^{JJ'}$ only with ${\hat A}_\mu^{*JJ'}$ and $J\ne J'$ in the planar approximation. Structure of vertices is determined by $d$ $M$-dimensional diagonal unitary matrices consisting of all classical absolute minima of zero action. Angles associated with these configurations differing only in group indices repel by Haar measure factors. At $V>1$ ${\hat A}_\mu$ inherits the constraints of ${\hat T}_\mu$ because of the factorized structure in terms of directions $\mu$. We still are ignoring the sparseness constraints.

	Based on equations (\ref{eqn:EK4},\ref{eqn:EK5},\ref{eqn:repl}) finite volume propagators are obtained from:
	\begin{equation}\label{eqn:finvolprop}		
	\begin{split}
		&\sum_{\mu, \nu} {\rm Tr}{\big [} T_\mu , T_\nu {\big ]} {\big [} T_\mu , T_\nu {\big ]}^\dagger  = \sum_{\mu, \nu} {\rm Tr}{\big [} {\hat T}_\mu , {\hat T}_\nu {\big ]} {\big [} {\hat T}_\mu , {\hat T}_\nu {\big ]}^\dagger  =
		\\&
		32 \sum_{\mu,J\ne J'} {\big |}{[{\hat A}_\mu ]}_{JJ'}{\big |}^2\sin^2 
		\frac{2\pi[(p_J)_\mu - (p_{J'})_\mu ] +[{\vartheta_{j_{J}}}^\mu-{\vartheta_{j_{J'}}}^\mu]}{2L_\mu} 
		\\&\times\sum_{\nu} \sin^2\frac{ 
		2\pi [(p_J)_\nu - (p_{J'})_\nu] + [{\vartheta_{j_{J}}}^\nu-{\vartheta_{j_{J'}}}^\nu]} {2L_\nu}.	
	\end{split} 
    \end{equation}
    The finite volume Faddeev-Popov determinant is:
    \begin{equation}
    	\prod_{J\ne J'} \sum_{\nu} \sin^2\frac{ 
    		2\pi [(p_J)_\nu - (p_{J'})_\nu] + [{\vartheta_{j_{J}}}^\nu-{\vartheta_{j_{J'}}}^\nu]}{2L_\nu}.
    \end{equation}
    Finally, we are ready to impose the sparseness constraints.
    The number of propagating complex fields in EKV at $V>1$, ignoring relative $1/N$ corrections, is $dVN^2/2$ 
    while, in the corresponding (same $M$) EK, it is $dV^2N^2/2$.
    The difference $dN^2V(V-1)/2$ counts the number of complex EK fields that need to be eliminated from EKV. The same number of fields
    have zero coefficient in equation (\ref{eqn:finvolprop}), due to the degeneracy. These non-propagating fields have $J\ne J'$ but equal $p_\mu$'s and angles. They differ only in their $p^{\perp(\mu)}$ 
    indices. This happens $L_\mu$ times for each arc on $S_\mu$ 
    and $V/L_\mu$ times per arc giving a factor of $V$. Recall that
    $(J,J')$ and $(J',J)$ entries in the sum are equal because
    only the absolute value of the field enters. So, there are  
    $dN^2(V^2-V)/2$ non-propagating fields, which leaves $dN^2 V/2$ propagating EK-fields, which is the right number for EKV. So, resurrecting the constraints ought to eliminate all non-propagating fields among the unconstrained ones, and nothing else. This indeed happens. 
		
	The $\mu$-propagators are explicitly dependent only on differences of $\mu$--indexed variables by equation (\ref{eqn:finvolprop}): the propagators already are explicitly circulant. In addition, for each $p_\mu$ there is a degeneracy factor with respect to the $p^{\perp(\mu)}$ indices, given by $V/L_\mu$. To eliminate the degeneracy by virtue of equation (\ref{unidir}) we set ${[{\hat A}_\mu]}_{(p^{\parallel \mu},(p^{\perp (\mu)};j)(p^{\parallel \mu},({p'}^{\perp (\mu)};j)}=0$ unless both perpendicular components vanish. This identifies each EKV planar diagram with a distinct EK one. 	
	
	In a nut-shell, instead of using double lines to only separate group index flows,like in continuum, at finite volume we use them to also. separate momentum index flows. Now, we can view the model as a pure matrix model where both momentum and group index summations are carried out by traces on combined indices. An injective map from EK to EKV degrees of freedom eliminates fields that became nonphysical as a consequence of clustering in the index bijection and reflected in  quadratic degeneracies.

	\section{Theorem.}
	{\bf To any order in planar lattice perturbation theory there is an exact equivalence among quenched versions of pure gauge} ${\mathbf U(N)}$  {\bf Wilson action on toroidal lattices of any finite size.}

    \subsection{Algorithm and Proof.}
    Take the perturbative expansion to some finite order. Keep only diagrams of planar topology because they will dominate as $N\;{\rm or}\;M\to\infty$ but do not take the limit yet. Feynman graphs of EK and EKV are identical with 
    vertices determined by the diagonal matrices corresponding to classical minima. In EKV all propagators are circulant in momenta.

    For any $V$ the angles associated with each direction $\mu$, appearing in absolute action minima and carrying only group indices are kept fixed to any order in the Feynman expansion. Expectation values of single trace operators made out of $T_\mu$-products are evaluated perturbatively at fixed angles and then averaged while fluctuation feedback on the distribution of angles is turned off. Observables of this nature can be included in the action with infinitesimal couplings, so we consider only vacuum diagrams.

     The series in bare lattice coupling constant should have a finite radius of convergence~\cite{Koplik:1977pf} so long as $N<\infty$. There is large $N$ volume independence, but the finite convergence radius  of the series may shrink to zero at $N=\infty$ since the volume effectively is infinite. This seems compatible with~
     \cite{tHooft:1982uvh,tHooft:1982ltl}.     
     
     Finally, the limit $N\to\infty$ is taken. To finite order in perturbation theory the 
     indices $p$ and $j$ can be replaced by continuum angles $\Theta^\mu$. 
     On each of the $L_\mu$ arcs on $S_\mu$ $\Theta^\mu$ monotonically and uniformly increases in intervals of $2\pi/N$ as $j=1,...N$. As $p_\mu=0,..,L_\mu-1$ consecutive arcs get sequentially added until the entire $2\pi$ range is covered. Equation (\ref{eqn:finvolprop})
     gets then rewritten as:
     \begin{equation}
      \prod_{\rho=1}^d \{ \int d\Theta^\rho\int d\Theta^{'\rho}\} \sum_\mu|[{\hat A}_\mu]_{\Theta,\Theta'}|^2 \sin^2\frac{\Theta^\mu-\Theta^{'\mu}}{2}
     \sum_\nu\sin^2\frac{\Theta^\nu-\Theta^{'\nu}}{2},
     \end{equation}
     and all $L_\mu$'s have disappeared. 
     
    \section{Summary and Comments.}

   All order planar Feynman diagrams of QEK type models exhibit large $N$ reduction. Taking $N\to\infty$ turns sums into integrals on the torus removing volume dependence. The UV cutoff stays finite in the limit. Momenta mesh with eigenvalue angles in DFT conjugated matrix structures representing dynamical variables.  Repulsion among Polyakov loop angles protects from IR divergences at $N<\infty$. 
   
   A general investigation of the angle distribution without quenching from the one loop effective potential for several $d$ and $L_\mu$ sets is currently in progress. 
   
   Perhaps the elementary approach here generalizes to twisted Eguchi-Kawai (TEK) model~\cite{Gonzalez-Arroyo:1982hwr}. Intuitively, TEK non-commutativity replaces flat space-time by a quantum phase-space, affording a more symmetrical treatment of momenta and sites. 
    
    Over the years, there have been several papers with titles mentioning ``orbifolding'', ``large $N$ reduction'' and ``volume independence'' in a continuum context. The author does not understand these articles well enough to comment on. In ~\cite{neubpoinc} the author employed orbifolding to prove large $N$ reduction in the Weingarten model~\cite{Weingarten:1979gn},\cite{Eguchi:1981kk} where the propagator is trivial. A crucial condition of the continuum field theory analysis in \cite{Bershadsky:1998cb} requires a factorized action of the orbifolding discrete group relative to total momenta flowing in propagators. This complicates matters. Fortunately, it turns out that the proof of volume independence without orbifolding given here is both instructive and simple.

    \section{Acknowledgments.}
    The sabbatical program of the Schwartz/Reisman Institute for Theoretical Physics at the Weizmann Institute of Science is acknowledged for support. 
    The high energy theory and phenomenology groups at Weizmann Institute are thanked for their friendly hosting.
    
\begin{singlespace}
	\bibliography{herbert10.bib}
\end{singlespace}	
\end{document}